\documentclass[showkeys,superscriptaddress,floatfix,prd,10pt,aps]{revtex4-2}
\usepackage{graphicx,epstopdf}
\usepackage[caption=false]{subfig}
\usepackage{appendix}
\usepackage[T1]{fontenc}
\usepackage{lmodern}
\usepackage[dvipsnames,x11names]{xcolor}
\usepackage[colorlinks=true,linkcolor=NavyBlue,citecolor=ForestGreen,urlcolor=NavyBlue]{hyperref}
\usepackage[sort&compress]{natbib}
\usepackage{lipsum}
\usepackage{morefloats}
\usepackage[pdf]{pstricks}
\usepackage{amsmath}
\usepackage{amssymb}
\usepackage{amsfonts}
\usepackage{rotating}
\usepackage{cancel}
\usepackage{mathtools}
\usepackage{bbm}
\usepackage{dsfont}
\usepackage{bbold}
\usepackage{multirow}
\usepackage{ulem}
\usepackage{physics}
\usepackage{orcidlink}
\usepackage{colortbl}

\def\xslash{x\!\!\!\slash }

\def\vel{\left|}
\def\ver{\right|}

\begin{document}

\title{Analysis of the $\mathrm{X_{AV}}$ state through its electromagnetic properties}

\author{Ula\c{s}~\"{O}zdem\orcidlink{0000-0002-1907-2894}
}%
\email[]{ulasozdem@aydin.edu.tr }
\affiliation{Health Services Vocational School of Higher Education, Istanbul Aydin University, Sefakoy-Kucukcekmece, 34295 Istanbul, T\"{u}rkiye}

\date{\today}
 
\begin{abstract}
To improve our understanding of the quark-gluon dynamics underlying multiquark states, we systematically study their electromagnetic properties. In this study, the magnetic and quadrupole moments of the theoretically predicted singly-charmed state with the quantum numbers $\mathrm{J^P = 1^+}$ is investigated within the framework of the QCD light-cone sum rules method by considering the diquark-antidiquark configuration of this state with quark contents $[ud][\bar{c}\bar{s}]$.  The predicted results for the magnetic and quadrupole moments are as $\mu_{\mathrm{X_{AV}}}=-0.89 ^{+0.14}_{-0.12}~\mu_N $ and $\mathcal{D}_{\mathrm{X_{AV}}} = (-0.46 ^{+0.07}_{-0.06})\times 10^{-2}  ~\mbox{fm}^2$. The results obtained can be useful in determining the exact nature of this state. This work will hopefully stimulate experimental interest in the study of the electromagnetic properties of multiquark systems. 
\end{abstract}
\keywords{Magnetic and quadrupole moments, tetraquarks, diquark-antidiquark picture,  QCD light-cone  sum rules}

\maketitle

\section{Motivation}
Besides the conventional hadron states, mesons, and baryons, it is theoretically possible in states containing more quarks.  
Since the discovery of the X(3872) state by the Belle Collaboration~\cite{Belle:2003nnu}, numerous hadronic states have been reported, that cannot be categorized in the traditional two- or three-quark configuration. After this discovery, the CDF, LHCb, Belle, CMS, BESIII, BaBar, and D0 collaborations also observed a large number of states, represented as XYZ states, pentaquarks, etc., which cannot be categorized in the traditional quark configuration. The discovery of these states has generated excitement in the scientific community and raised questions about their exact nature and internal structure. Numerous models have been proposed to clarify and determine the nature of these states, and extensive research is currently being conducted on them. The observation of the above-mentioned states triggered interesting theoretical studies of these new states in the context of different models and approaches aimed at revealing their nature, quantum numbers, and internal structure. There are many excellent reviews on this topic that can be found in the literature~\cite{Faccini:2012pj,Esposito:2014rxa,Chen:2016qju,Ali:2017jda,Esposito:2016noz,Olsen:2017bmm,Lebed:2016hpi,Guo:2017jvc,Nielsen:2009uh,Brambilla:2019esw,Liu:2019zoy, Agaev:2020zad, Dong:2021juy,Chen:2015ata,Meng:2022ozq,Chen:2022asf}.

In 2021, two exotic states with minimal quark contents $[\bar{c}\bar{s}][ud]$, were observed by LHCb Collaboration using full amplitude analysis $B^+\to D^+D^-K^+$ decays~\cite{LHCb:2020bls,LHCb:2020pxc}.  The masses and widths of these states are
measured as
\begin{align}
\mathrm{X_0(2900)}: M&=2866\pm7\pm2~\text{MeV},~~~~
\Gamma=57\pm12\pm4~\text{MeV},\nonumber\\
\mathrm{X_1(2900)}: M&=2904\pm5\pm1~\text{MeV},~~~~
\Gamma=110\pm11\pm4~\text{MeV}.\nonumber
\end{align}

The LHCb Collaboration predicted that $\mathrm{X_0(2900)}$ and $\mathrm{X_1(2900)}$ have the quantum numbers $\mathrm{J^P = 0^+}$ and $\mathrm{1^-}$, respectively. In various combinations, $c$, $s$, $u$, and $d$ quarks form different categories of exotic states, properties of which deserve further investigation.  In Refs. \cite{Molina:2020hde,Dai:2022qwh,Dai:2022htx,Sundu:2022kyd,Azizi:2021aib,Agaev:2021jsz,Azizi:2018mte,Agaev:2017oay}, apart from $\mathrm{J^P = 0^+}$ and $\mathrm{1^-}$ states, two other states $\mathrm{J^P = 1^+}$ ($\mathrm{X_{AV}}$) and $\mathrm{2^+}$ were also predicted, which are not yet observed. 
To understand the exact nature and internal structure of these states, it is useful to study their spectroscopic parameters, as well as other characteristics such as electromagnetic and weak decays.

The electromagnetic properties, especially the magnetic and quadrupole moments, are prominent observables of the hadrons that can be calculated and measured in the same way as the mass and the decay. The magnetic and quadrupole moments are of particular interest in studying the inner structure and possible deformation of hadrons.   Furthermore, the magnetic moment of a hadron is a measure of its ability to interact with magnetic fields, making it a crucial parameter for understanding hadron behavior. Inspired by this, in the present work, the magnetic and quadrupole moments of the theoretically predicted singly-charmed state, $\mathrm{X_{AV}}$, with the quantum numbers $\mathrm{J^P = 1^+}$ are extracted by means of the QCD light-cone sum rules method by considering the diquark-antidiquark configuration of this state with quark contents $[ud][\bar{c}\bar{s}]$.  Several studies in the literature have extracted the magnetic and quadrupole moments of hidden-charm and singly-charmed tetraquark states~\cite{Ozdem:2021yvo,Ozdem:2017jqh,Xu:2020qtg,Wang:2017dce,Xu:2020evn,Ozdem:2022kck,Ozdem:2021hka,Ozdem:2023frj,Wang:2023vtx, Azizi:2021aib,Ozdem:2022ydv,Azizi:2018mte}.
 
The outline of the article is as follows.  In Sec. \ref{secII} we present the detailed QCD light-cone sum rule calculations for the $\mathrm{X_{AV}}$ state. In Sec. \ref{secIII} we performed a numerical analysis of the success of the obtained sum rules and extracted the numerical values of the magnetic and quadrupole moments of the $\mathrm{X_{AV}}$ state.  The results obtained are summarized in Sec \ref{secIV}.

\begin{widetext}

\section{ Theoretical framework}\label{secII}

The QCD light-cone sum rules method is a powerful tool for exploring conventional and exotic hadron characteristics and has been widely used to extract the spectroscopic parameters, magnetic and quadrupole moments, form factors, etc. of such states. In this method, the correlation function, which is the key component of the method, is calculated both regarding hadronic features (the hadronic representation) and the QCD features (the QCD representation) based on the prescription of the method.  Then, by equating these two different representations of the correlation function, the physical quantities to be calculated are obtained~\cite{Chernyak:1990ag, Braun:1988qv, Balitsky:1989ry}.

To determine the magnetic and quadrupole moments of the $\mathrm{X_{AV}}$ state, we begin by analyzing the following correlation function: 
\begin{equation}
 \label{edmn01}
\Pi _{\mu \nu }(p,q)=i\int d^{4}xe^{ip\cdot x}\langle 0|\mathcal{T}\{J_{\mu}(x)
J_{\nu }^{\dagger }(0)\}|0\rangle_{\gamma}, 
\end{equation}
where q is the photon's momentum and the subindex $\gamma$ is the weak external electromagnetic background field. Here $J_{\mu}(x)$ is the interpolating current of the $\mathrm{X_{AV}}$ state, which is given as 
\begin{align}
\label{curr}
    J_{\mu}(x)&=\epsilon \tilde{\epsilon}\big\{
    [ { u^{bT}}(x) C \gamma_5  d^c(x)]
    [ \bar c^d(x) \gamma_{\mu} C { \bar s^{eT}}(x)] 
\big\},
  \end{align}
where  $\epsilon =\epsilon ^{abc}$; $\tilde{\epsilon}=\epsilon ^{ade}$; the $a$, $b$, $c$, $d$, and  $e$  are color indices;  and $C$ denotes the charge conjugation operator.

As a first step in the analysis, let us calculate the hadronic representation of the correlation function.  To obtain the hadronic representation of the desired sum rules, complete sets of hadronic states with the same quantum numbers as the considered hadrons are inserted into the correlation function. As a result, we get
 \begin{align}
\label{edmn04}
\Pi_{\mu\nu}^{Had} (p,q) &= \frac{\langle 0 \mid J_\mu (x) \mid \mathrm{X_{AV}}(p, \varepsilon^i) \rangle }{ m_{\mathrm{X_{AV}}}^2 - p^2}   
\langle \mathrm{X_{AV}}(p, \varepsilon^i) \mid \mathrm{X_{AV}}(p+q, \varepsilon^f) \rangle_\gamma  
\frac{\langle \mathrm{X_{AV}}(p+q,\varepsilon^f) \mid {J^\dagger}_\nu (0) \mid 0 \rangle }{m_{\mathrm{X_{AV}}}^2 - (p+q)^2}
+\mbox{higher states},
\end{align}
where the $\varepsilon^{i}$ and $\varepsilon^{f}$ are the polarization vectors of initial and final $\mathrm{X_{AV}}$ state, respectively.  The matrix elements in  Eq. (\ref{edmn04}) are required for further calculations, which are given as
\begin{align}
\langle 0 \mid J_\mu(x) \mid \mathrm{X_{AV}}(p,\varepsilon^i) \rangle &= \lambda_{\mathrm{X_{AV}}} \varepsilon_\mu^i\,,\\
\langle \mathrm{X_{AV}}(p+q,\varepsilon^f) \mid {J^\dagger}_\nu (0) \mid 0 \rangle  &= \lambda_{\mathrm{X_{AV}}} \varepsilon_\nu^{*f}\,, \\
\langle \mathrm{X_{AV}}(p,\varepsilon^i) \mid  \mathrm{X_{AV}} (p+q,\varepsilon^{f})\rangle_\gamma &= - \varepsilon^\gamma (\varepsilon^{i})^\alpha (\varepsilon^{f})^\beta \Big\{ 
  G_1(Q^2) (2p+q)_\gamma ~g_{\alpha\beta} 
+ G_2(Q^2) ( g_{\gamma\beta}~ q_\alpha -  g_{\gamma\alpha}~ q_\beta) \nonumber\\
&
- \frac{1}{2 m_{\mathrm{X_{AV}}}^2} G_3(Q^2)~ (2p+q)_\gamma  
q_\alpha q_\beta  \Big\},\label{edmn06}
\end{align}
where $\lambda_{\mathrm{X_{AV}}}$ being the current coupling constant of the $\mathrm{X_{AV}}$ state, $\varepsilon^\gamma$ is the polarization vector of the photon, and  $G_1(Q^2)$,  $G_2(Q^2)$, and $G_3(Q^2)$ are Lorentz invariant form factors with $Q^2=-q^2$.

The final form of the hadronic part of the correlation function is obtained by using the Eqs.~(\ref{edmn04})-(\ref{edmn06}) in the following manner:
%
%
\begin{align}
\label{edmn09}
 \Pi_{\mu\nu}^{Had}(p,q) &=  \frac{\varepsilon^\tau \, \lambda_{\mathrm{X_{AV}}}^2}{ [m_{\mathrm{X_{AV}}}^2 - (p+q)^2][m_{\mathrm{X_{AV}}}^2 - p^2]}
 \bigg\{G_1(Q^2)  (2p+q)_\tau\bigg(g_{\mu\nu}-\frac{p_\mu p_\nu}{m_{\mathrm{X_{AV}}}^2}
 -\frac{(p+q)_\mu (p+q)_\nu}{m_{\mathrm{X_{AV}}}^2} \nonumber\\
 &+\frac{(p+q)_\mu p_\nu}{2m_{\mathrm{X_{AV}}}^4} (Q^2+2m_{\mathrm{X_{AV}}}^2)
 \bigg)
 + G_2 (Q^2) \bigg(q_\mu g_{\tau\nu} - q_\nu g_{\tau\mu} -
\frac{p_\nu}{m_{\mathrm{X_{AV}}}^2}  \big(q_\mu p_\tau  - \frac{1}{2}
Q^2 g_{\mu\tau}\big) \nonumber\\
& 
+
\frac{(p+q)_\mu}{m_{\mathrm{X_{AV}}}^2}  \big(q_\nu (p+q)_\tau+ \frac{1}{2}
Q^2 g_{\nu\tau}\big)-  
\frac{(p+q)_\mu p_\nu p_\tau}{m_{\mathrm{X_{AV}}}^4} \, Q^2
\bigg)
\nonumber\\
&
-\frac{G_3(Q^2)}{m_{\mathrm{X_{AV}}}^2}(2p+q)_\tau \bigg(
q_\mu q_\nu -\frac{p_\mu q_\nu}{2 m_{\mathrm{X_{AV}}}^2} Q^2 +\frac{(p+q)_\mu q_\nu}{2 m_{\mathrm{X_{AV}}}^2} Q^2
-\frac{(p+q)_\mu q_\nu}{4 m_{\mathrm{X_{AV}}}^4} Q^4\bigg)
\bigg\}\,.
\end{align}

Since the magnetic and quadrupole moments are related to the magnetic ($F_M(Q^2)$) and quadrupole ($F_{\mathcal D}(Q^2)$) form factors, these form factors need to be written in terms of the $G_1(Q^2)$, $G_2(Q^2)$  and $G_3(Q^2)$  form factors, which are given as follows 
\begin{align}
\label{edmn07}
&F_M(Q^2) = G_2(Q^2)\,,\nonumber \\
&F_{\mathcal D}(Q^2) = G_1(Q^2)-G_2(Q^2)+\Big(1+\frac{Q^2}{4 m_{\mathrm{X_{AV}}}^2}\Big) G_3(Q^2)\,,
\end{align}
At zero momentum square, $Q^2=0$,  the magnetic and quadrupole form factors are equal to the magnetic ($\mu_{\mathrm{X_{AV}}}$) and quadrupole ($\mathcal {D}_{\mathrm{X_{AV}}}$) moments.  In this limit, the $\mu_{\mathrm{X_{AV}}}$ and $\mathcal {D}_{\mathrm{X_{AV}}}$ can be written in terms of the $F_M(0)$ and $F_{\mathcal D}(0)$ factors as follows,
\begin{align}
\label{edmn08}
& \mu_{\mathrm{X_{AV}}}  = \frac{e}{2 m_{\mathrm{X_{AV}}} }F_M(0) \,, \nonumber\\
& {\mathcal {D}_{\mathrm{X_{AV}}}}  =\frac{e}{ m_{\mathrm{X_{AV}}}^2} F_{\cal D}(0)\,.
\end{align}

We are now ready to start calculating the QCD representation of the correlation function. In the QCD representation, we contract all of the quark fields in the correlation function with Wick’s theorem. According to the procedures described above, the QCD representation of the correlation function for the $\mathrm{X_{AV}}$ state is obtained as follows 

\begin{align}
\label{neweq2}
\Pi _{\mu \nu }^{\mathrm{QCD-\mathrm{X_{AV}}}}(p,q)&=i \epsilon \tilde{\epsilon} \epsilon^{\prime} \tilde{\epsilon}^{\prime}
\int d^{4}xe^{ipx} \langle 0 | \Big\{ 
\mathrm{Tr}\Big[ \gamma _{5 }{\widetilde S}_{u}^{bb^{\prime }}(x) \gamma _{5 }S_{d}^{cc^{\prime }}(x)\Big] 
\mathrm{Tr}\Big[\gamma_{\mu}{\widetilde S}_{s}^{e^{\prime }e}(-x)\gamma_{\nu} S_{c}^{d^{\prime }d}(-x)\Big] 
 \Big\}| 0 \rangle_\gamma,
\end{align} 
%
where $\epsilon =\epsilon ^{abc}$,  $\tilde{\epsilon}=\epsilon ^{ade}$, $\epsilon^{\prime } =\epsilon ^{a^{\prime }b^{\prime }c^{\prime }}$, and  $\tilde{\epsilon^{\prime }}=\epsilon ^{a^{\prime }d^{\prime }e^{\prime }}$;  $S_{c}(x)$ and $S_{q}(x)$ represent propagators of  heavy and light quarks. Here we also use the notation ${\widetilde S}_{q(c)}(x) = C S_{q(c)}^T (x) C$.  The propagators of light and heavy quarks are given by the following formula~\cite{Yang:1993bp, Belyaev:1985wza}:
%
 %
\begin{align}
\label{edmn12}
S_{q}(x) &= S_q^{free}(x)
- \frac{\langle \bar qq \rangle }{12} \Big(1-i\frac{m_{q} \xslash}{4}   \Big)
- \frac{\langle \bar q \sigma.G q \rangle }{192}x^2  \Big(1 -i\frac{m_{q} \xslash}{6}   \Big)
-\frac {i g_s }{32 \pi^2 x^2} ~G^{\mu \nu} (x) \Big[\rlap/{x}
\sigma_{\mu \nu} +  \sigma_{\mu \nu} \rlap/{x}
 \Big],
 \\
S_{c}(x)&=S_c^{free}(x)
-\frac{g_{s}m_{c}}{16\pi ^{2}} \int_0^1 dv\, G^{\mu \nu }(vx)\bigg[ (\sigma _{\mu \nu }{\xslash}
  +{\xslash}\sigma _{\mu \nu })
  \frac{K_{1}\Big( m_{c}\sqrt{-x^{2}}\Big) }{\sqrt{-x^{2}}}
+2\sigma_{\mu \nu }K_{0}\Big( m_{c}\sqrt{-x^{2}}\Big)\bigg],
\label{edmn13}
\end{align}%
%
where 
\begin{align}
S_q^{free}(x) &=\frac{1}{2 \pi^2 x^2}\Big( i \frac{{\xslash}}{x^{2}}-\frac{m_{q}}{2 } \Big),\\
S_c^{free}(x) &= \frac{m_{c}^{2}}{4 \pi^{2}} \bigg[ \frac{K_{1}\Big(m_{c}\sqrt{-x^{2}}\Big) }{\sqrt{-x^{2}}}
+i\frac{{\xslash}~K_{2}\Big( m_{c}\sqrt{-x^{2}}\Big)}
{(\sqrt{-x^{2}})^{2}}\bigg].
\end{align} 

The QCD representation of the correlation function includes two different contributions that need to be calculated: perturbative (short-distance) and non-perturbative (long-distance).  To calculate the short-distance contribution, it is sufficient to substitute one of the light/heavy-quark propagators in Eq.~(\ref{neweq2}) in the following way
\begin{align}
\label{free}
S^{free}(x) \rightarrow \int d^4y\, S^{free} (x-y)\,\rlap/{\!A}(y)\, S^{free} (y)\,,
\end{align}
where the rest of the propagators are taken into account as free propagators. This amounts to taking $\bar T_4^{\gamma} (\underline{\alpha}) = 0$ and $S_{\gamma} (\underline {\alpha}) = \delta(\alpha_{\bar q})\delta(\alpha_{q})$ as the light-cone distribution amplitude in the three particle distribution amplitudes (see Ref. \cite{Li:2020rcg}). 

To obtain the long-distance contributions, it is sufficient to replace one of the light-quark propagators in Eq.(\ref{neweq2}) with the following expression
\begin{align}
\label{edmn14}
S_{\mu\nu}^{ab}(x) \rightarrow -\frac{1}{4} \big[\bar{q}^a(x) \Gamma_i q^b(0)\big]\big(\Gamma_i\big)_{\mu\nu},
\end{align}
 where   $\Gamma_i = \mathrm{1}, \gamma_5, \gamma_\mu, i\gamma_5 \gamma_\mu, \sigma_{\mu\nu}/2$. 
 
 After the aforementioned light-quark replacement, the rest of the propagators are considered to be full propagators, including both perturbative and non-perturbative contributions. The matrix elements of nonlocal operators such as $\langle \gamma(q)\vel \bar{q}(x) \Gamma_i G_{\mu\nu}q(0) \ver 0\rangle$  and $\langle \gamma(q)\vel \bar{q}(x) \Gamma_i q(0) \ver 0\rangle$, which are expressed regarding photon distribution amplitudes (DAs), come out when a photon interacts non-perturbatively with light-quark fields (for details see Ref. \cite{Ball:2002ps}).  These steps are standard for this method and are quite lengthy, so we do not present them in the text. For interested readers, details of this procedure performed to acquire the expression of the perturbative and non-perturbative contributions are presented in Refs.~\cite{Ozdem:2022kck,Ozdem:2022eds}.  Using the above-mentioned procedures and then applying the Fourier transform to the obtained expressions to transfer the position space expressions to the momentum space, the QCD representation of the correlation function is extracted. It should be noted that the photon DAs utilized in this study only take into account contributions from light quarks. However, in principle, the photon can be emitted at a long-distance from the charm quark.  In technical terms, the matrix elements of nonlocal operators are proportional to the product of DAs, quark condensates, and some non-perturbative constants.   Knowing that the contribution of non-perturbative constants to our analysis is negligible even in the case of light quarks, we can neglect them in the case of heavy quarks. The heavy-quark condensates are known to be proportional to $1/m_Q$. Due to the large mass of the heavy quarks, such condensates are largely suppressed for the heavy quarks~\cite{Antonov:2012ud}.  Thus, our computations excluded DAs containing heavy quarks, which are long-distance contributions. We only considered the short-distance photon emission from the heavy quarks, as described in  Eq.~(\ref{free}).

%
By separating the coefficients of structures $(\varepsilon.p) (p_\mu q_\nu -p_\nu q_\mu) $ and $(\varepsilon.p) q_\mu q_\nu$ from the QCD and hadronic sides of the correlation function and equating them, we can determine the magnetic and quadrupole moments of the $\rm{X_{AV}}$ state.  To suppress the contributions of the higher states and continuum, we perform double Borel transformation on the variables $p^2$ and $(p+q)^2$, and continuum subtraction. Note that Borel transformations are performed by means of the equations
\begin{align}
 \mathcal{B}\bigg\{ \frac{1}{\big[ [p^2-m^2_i][(p+q)^2-m_f^2] \big]}\bigg\} \rightarrow e^{-m_i^2/M_1^2-m_f^2/M_2^2}
\end{align}
in the hadronic side, and 
\begin{align}
 \mathcal{B}\bigg\{ \frac{1}{\big(m^2- \bar u p^2-u(p+q)^2\big)^{\alpha}}\bigg\} \rightarrow (M^2)^{(2-\alpha)} \delta (u-u_0)e^{-m^2/M^2},
\end{align}
 in the QCD side,  where we use
\begin{align*}
 {M^2}= \frac{M_1^2 M_2^2}{M_1^2+M_2^2}, ~~~
 u_0= \frac{M_1^2}{M_1^2+M_2^2}.
\end{align*}
  Here $ M_1^2 $ and $ M_2^2 $ being the Borel parameters in the initial and final states, respectively. Since we have the same $\rm{X_{AV}}$ in the initial and final states, therefore we can set, M$_1^2$ = M$_2^2$= 2M$^2$ and $u_0 = 1/2 $, which leads to the our approximation being sufficient to suppress higher states and continuum contributions. As the result of these computations, we get the following sum rules for the magnetic and quadrupole moments of the $\rm{X_{AV}}$ state,
\begin{align}
 \label{sonj1}
  \mu_{\mathrm{X_{AV}}} \lambda_{\mathrm{X_{AV}}}^2 &= e^{\frac{m_{\mathrm{X_{AV}}}^2}{\mathrm{M^2}}} \,\, \Delta_1^{\mathrm{QCD}}(\mathrm{M^2},\mathrm{s_0}),\\
  \mathcal{D}_{\mathrm{X_{AV}}} \lambda_{\mathrm{X_{AV}}}^2 &= m_{\mathrm{X_{AV}}}^2 e^{\frac{m_{\mathrm{X_{AV}}}^2}{\mathrm{M^2}}}  \,\, \Delta_2^{\mathrm{QCD}}(\mathrm{M^2},\mathrm{s_0}).\label{sonj2}
 \end{align}

 For brevity, only the explicit expressions of the $\Delta_1^{\mathrm{QCD}}(\mathrm{M^2},\mathrm{s_0})$ function are listed in the text, since the $\Delta_2^{\mathrm{QCD}}(\mathrm{M^2},\mathrm{s_0})$ function is similar in form. 
\begin{align}
\Delta_1^{\mathrm{QCD}}(\mathrm{M^2},\mathrm{s_0})&= \frac {1} {1536 \pi^6}\bigg[
   m_c^8\Big ((e_c + 4 e_s) m_c^4 I[-3] + 
       4 (e_c + 3 e_s) m_c^2 I[-2] + 6 (e_c + 2 e_s) I[-1]\Big) + 
    4 (e_c + e_s) m_c^6 I[0] \nonumber\\
    &+ e_c m_c^4 I[1] + 
    16 (e_c + 2 e_s) I[3] \bigg]\nonumber\\
     &+\frac{m_c \langle g_s^2 G^2\rangle \langle \bar ss \rangle f_{3\gamma}}{5184  \pi^2} (e_u+e_d)  I[-2] \psi^
   a[u_ 0]\nonumber\\
   &+\frac { \langle \bar qq \rangle^2 (e_d + 
     e_u)} {192 m_c^2 \pi^2}\Big (m_ 0^2 m_c^4 I[-2] + 
   4 m_c^2 I[0] - 2 I[1]\Big) I_ 3[\mathcal S] \nonumber
   \end{align}
         \begin{align}
   &+\frac {m_c \langle g_s^2 G^2\rangle \langle \bar ss \rangle} {165888 \pi^4}\bigg[   -33 e_s \Big (m_c^2 \
I[-2] - I[-1]\Big) I_ 4[\mathcal S] - 
    33 e_s \Big (m_c^2 I[-2] - I[-1]\Big) I_ 4[\mathcal {\tilde S}] - 
    288 e_s m_c^2 I_ 6[
         h_ {\gamma}] \nonumber\\
         &\times I[-2]\bigg]\nonumber\\
     &-\frac {\langle g_s^2 G^2\rangle  f_{3\gamma} } {13824 m_c^2 \pi^4}\bigg[
   2 m_c^2 \bigg ((e_d + e_u) m_c m_s \Big (m_c^2 I[-2] - 
          I[-1]\Big) - (e_d - 6 e_s + e_u) I[0]\bigg) + (e_d - 6 e_s +
        e_u) I[1]  \bigg]\nonumber\\
          & \times\psi^a[u_ 0]\nonumber\\ 
   &-\frac {m_c \langle \bar ss \rangle} {1152 \pi^4}\bigg[   (e_d + 
       e_u) f_ {3\gamma} m_ 0^2 m_c m_s \pi^2 I_ 2[\mathcal V] I[-2] \
+ 144 e_s \Big (m_c^8 I[-3] + 2 m_c^6 I[-2] + m_c^4 I[-1] + 
       4 I[1]\Big) \nonumber\\
       & \times I_ 6[h_ {\gamma}] \bigg]\nonumber\\
  &+\frac {f_ {3\gamma}  m_c^4} {3072 \pi^4}\bigg[-(e_d + e_u)\Big (5 m_c^6 I[-3] - 12 m_c^4 I[-2] + 9 m_c^2 I[-1] - 
    2 I[0]\Big) I_ 2[\mathcal V] + 
 48 e_s m_c^2 \Big (m_c^4 I[-3] \nonumber\\
 &- 2 m_c^2 I[-2] + I[-1]\Big) \psi^
    a[u_0]\bigg],
     \end{align}
where $\langle \bar qq \rangle$, $\langle \bar ss \rangle$ and $\langle g_s^2 G^2\rangle$  are u/d-quark,  s-quark and gluon condensates, respectively. The functions $I[n]$, $I_1[\mathcal{A}]$, $I_2[\mathcal{A}]$, $I_3[\mathcal{A}]$, $I_4[\mathcal{A}]$, $I_5[\mathcal{A}]$ and $I_6[\mathcal{A}]$    are defined as
\begin{align}
I[n]&= \int_{m_c^2}^{s_0} ds \,s^n e^{-s/M^2}\nonumber\\
 I_1[\mathcal{A}]&=\int D_{\alpha_i} \int_0^1 dv~ \mathcal{A}(\alpha_{\bar q},\alpha_q,\alpha_g)
 \delta'(\alpha_ q +\bar v \alpha_g-u_0),\nonumber\\
   I_2[\mathcal{A}]&=\int D_{\alpha_i} \int_0^1 dv~ \mathcal{A}(\alpha_{\bar q},\alpha_q,\alpha_g)
 \delta'(\alpha_{\bar q}+ v \alpha_g-u_0),\nonumber\\
   I_3[\mathcal{A}]&=\int_0^1 du~ A(u)\delta'(u-u_0),\nonumber\\
  I_4[\mathcal{A}]&=\int D_{\alpha_i} \int_0^1 dv~ \mathcal{A}(\alpha_{\bar q},\alpha_q,\alpha_g)
 \delta(\alpha_ q +\bar v \alpha_g-u_0),\nonumber\\
   I_5[\mathcal{A}]&=\int D_{\alpha_i} \int_0^1 dv~ \mathcal{A}(\alpha_{\bar q},\alpha_q,\alpha_g)
 \delta(\alpha_{\bar q}+ v \alpha_g-u_0),\nonumber\\
 I_6[\mathcal{A}]&=\int_0^1 du~ A(u),\nonumber
 \end{align}
where $\mathcal{A}$ stands for the corresponding photon DAs. Here the ${\cal D} \alpha_i$ can be written as
 
\begin{eqnarray}
\label{nolabel05}
\int {\cal D} \alpha_i = \int_0^1 d \alpha_{\bar q} \int_0^1 d
\alpha_q \int_0^1 d \alpha_g \,  \delta(1-\alpha_{\bar
q}-\alpha_q-\alpha_g)~.
\end{eqnarray}
\end{widetext}

\section{Numerical results and discussions} \label{secIII}
 
This section presents numerical analyses for the magnetic and quadrupole moments of the $\mathrm{X_{AV}}$ state.  The parameters that are used in our calculations are as follows: $m_u=m_d=0$, $m_s =96^{+8.0}_{-4.0}\,\mbox{MeV}$,
$m_c = (1.27\pm 0.02)\,$GeV, 
 $m_{\mathrm{X_{AV}}}= 2800 \pm 75$~MeV \cite{Sundu:2022kyd}, $\langle \bar ss\rangle $= $0.8 \langle \bar qq\rangle$ with 
$\langle \bar qq\rangle $ =$(-0.24\pm0.01)^3\,$GeV$^3$  \cite{Ioffe:2005ym},   $m_0^{2} = 0.8 \pm 0.1$~GeV$^2$, 
$\langle g_s^2G^2\rangle = 0.88~ $GeV$^4$~\cite{Matheus:2006xi}, and $f_{3\gamma}=-0.0039~$GeV$^2$~\cite{Ball:2002ps}. From Equation (\ref{sonj1})-(\ref{sonj2}) it is concluded that the current coupling constant of the $\mathrm{X_{AV}}$ state is required to calculate the magnetic and quadrupole moments. The current coupling constant of this state is calculated in Ref.~\cite{Sundu:2022kyd} and we will use this value in our numerical analysis.  One of the important ingredients of the QCD light-cone sum rules for the magnetic and quadrupole moments is the photon DAs. The expressions of the DAs are borrowed from Ref. \cite{Ball:2002ps}.

 In addition to the aforementioned input parameters,  there are two free parameters in Eqs. (\ref{sonj1})-(\ref{sonj2}): the Borel mass $\mathrm{M^2}$ and the continuum threshold parameter $\mathrm{s_0}$. To obtain reliable QCD sum rule results, we require that the $\mathrm{s_0}$ dependence and the $\mathrm{M^2}$ dependence of the magnetic and quadrupole moment predictions should be weak. To achieve this, we need to carefully examine the convergence of the operator product expansion (OPE)  and the pole contribution (PC). The convergence of the OPE is generally required to be sufficiently small to ensure convergence of the operator product expansion series i.e., convergence of OPE should be under control, and the PC needs to be as large as possible to ensure the effectiveness of the single-pole approach.   The formulas below can be used for the definition of these constraints:
\begin{align}
 \mbox{PC} &=\frac{\Delta (\mathrm{M^2},\mathrm{s_0})}{\Delta (\mathrm{M^2},\infty)} \geq  30\%,\\
 \nonumber\\
 \mathrm{R(M^2)} &=\frac{\Delta^{\mathrm{DimN}} (\mathrm{M^2},\mathrm{s_0})}{\Delta (\mathrm{M^2},\mathrm{s_0})}\leq  5\%,
 \end{align}
 where $\Delta^{\mathrm{DimN}} (\mathrm{M^2},\mathrm{s_0})$ is a sum of $\mbox{DimN} = \mathrm{Dim(7+8+9)}$.   After applying the above formulas, the obtained PC, convergence of OPE, and working intervals of the $\mathrm{M^2}$ and $\mathrm{s_0}$ are presented in Table \ref{parameter}. It follows that the working intervals determined for $\mathrm{M^2}$ and $\mathrm{s_0}$ satisfy the limitations imposed by the PC and the convergence of OPE. For completeness, the extracted predictions for the magnetic and quadrupole moments of the  $\mathrm{X_{AV}}$ state are shown in Fig. 1.  From this figure, it can be seen that the magnetic and quadrupole moments of this state show good stability concerning the variation of $\mathrm{M^2}$ in its working interval.
\begin{widetext}

\begin{table}[htb!]
	\addtolength{\tabcolsep}{10pt}
	\caption{
	The Borel windows, continuum threshold parameters, pole contributions, and convergence of the OPE for the magnetic and quadrupole moments of the  $\mathrm{X_{AV}}$ state.
	}
	\label{parameter}
		\begin{center}
\begin{tabular}{l|ccccc}
                \hline\hline
                \\
~~~~~State~~ & $\mathrm{s_0}$ (GeV$^2$)& 
$\mathrm{M^2}$ (GeV$^2$) & ~~  PC ($\%$) ~~ &   $\mathrm{R(M^2)}$ 
 ($\%$) \\
\\
                                        \hline\hline
                                        \\
   ~~   $\mathrm{X_{AV}}$ ~~&  $[10.5, 12.0]$ & $[2.8, 3.4]$ & $[53, 37]$ & $ 3.73$   \\
   \\
                                       \hline\hline
 \end{tabular}
\end{center}
\end{table}

\end{widetext}

%
\begin{figure}[htb!]
\centering
 \includegraphics[width=0.45\textwidth]{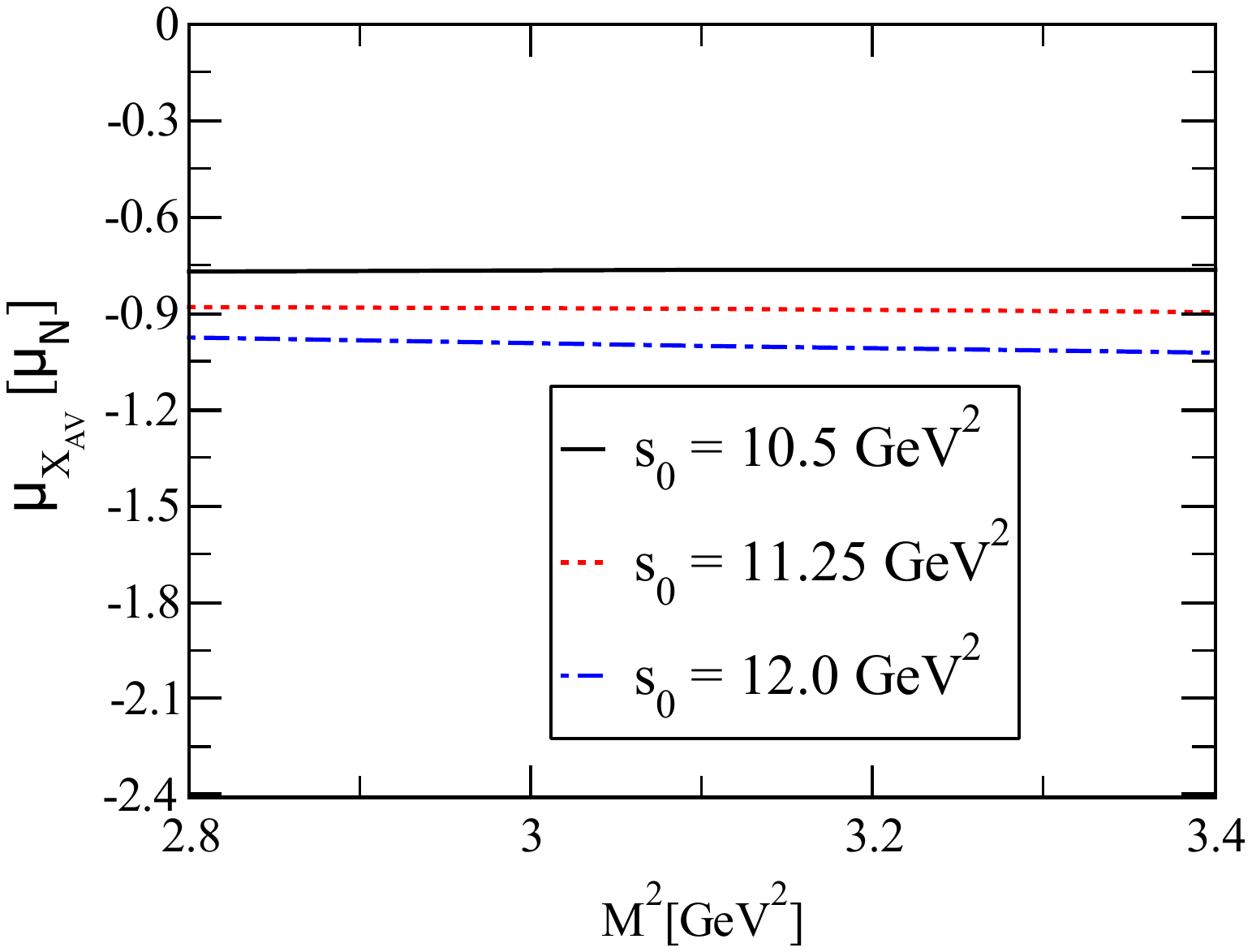}~~~~
 \includegraphics[width=0.45\textwidth]{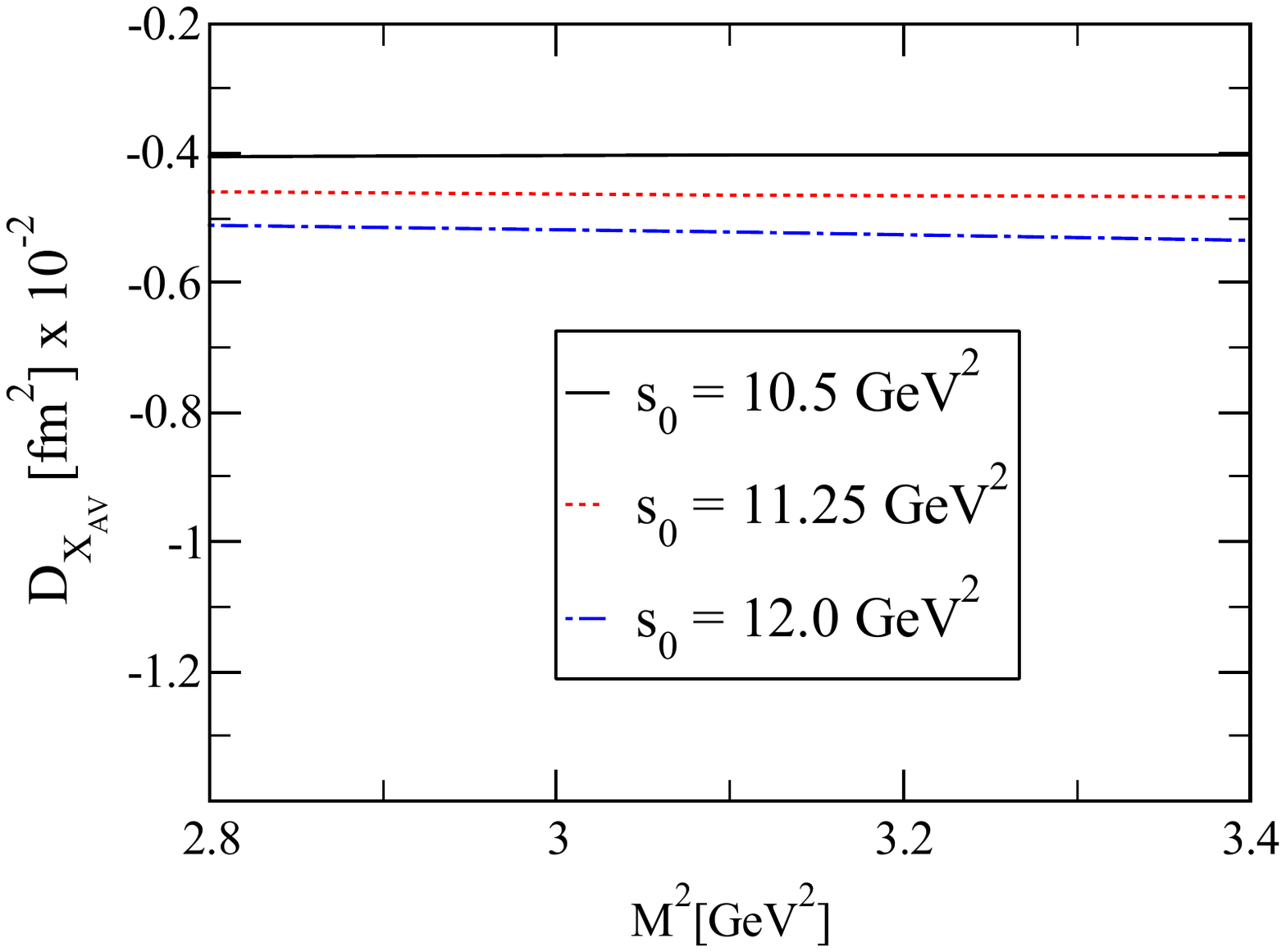}\\
 \caption{Dependence of the magnetic and quadrupole moments of the $\mathrm{X_{AV}}$ on $\mathrm{M^2}$ at fixed values of the continuum threshold $\mathrm{s_0}$.}
 \label{Msqfig}
  \end{figure}
%

%
The magnetic and quadrupole moments of the $\mathrm{X_{AV}}$ state are determined by calculating them at different  $\mathrm{M^2}$ and $\mathrm{s_0}$ from the interval given in Table~\ref{parameter} and averaging the obtained results to the find mean values of these parameters.  The final results for magnetic and quadrupole moments are presented as follows
\begin{align}
 \mu_{\mathrm{X_{AV}}}& = -0.89 ^{+0.14}_{-0.12}~\mu_N, \\
 \nonumber\\
 \mathcal{D}_{\mathrm{X_{AV}}} &= (-0.46 ^{+0.07}_{-0.06})\times 10^{-2} ~\mbox{fm}^2.
\end{align}

The uncertainties in the results are due to the variation of the parameters $\mathrm{M^2}$, $\mathrm{ s_0}$, and the errors in the values of the input parameters. By examining the magnetic moment result, we can assume that the magnetic moment of the  $\mathrm{X_{AV}}$ state is sufficiently large to be measured in future experiments. In the case of the quadrupole moment, we obtain a non-zero but small value, indicating a non-spherical charge distribution.  The predicted sign of the quadrupole moment allows us to say that the $\mathrm{X_{AV}}$ state has oblate charge distributions.  Since there are no experimental or theoretical predictions in the literature that we can compare, perhaps the reader will have a better idea of the magnetic moment results obtained if we compare the $\mathrm{X_{AV}}$ state with the result of the $\mathrm{X_1(2900)}$ state. In Ref.~\cite{Ozdem:2022ydv}, the magnetic moment of the  $\mathrm{X_1(2900)}$ state was extracted through the QCD light-cone sum rules by assuming that the $\mathrm{X_1(2900)}$ state is considered as the diquark-antidiquark picture.  The obtained result is  as $\mu_{\mathrm{X_1(2900)}}= 0.79^{+0.36}_{-0.39}~\mu_N$. 
 As you can see, the absolute value of the result that we obtained is in the same order of magnitude as the state that we have mentioned.

To gain a deeper understanding of the underlying quark-gluon dynamics, it is useful to consider the contributions made by the individual quark sectors to the magnetic and quadrupole moments. When this has been done, we obtain that the terms proportional to the $e_c$ contribute about 27$\%$ to the total results, $e_s$ is about 33$\%$, $e_u$ is about 27$\%$, and $e_d$ about 13$\%$.

\section{Summary and outlook}\label{secIV}

We systematically study the electromagnetic properties of multiquark systems to improve our understanding of the underlying quark-gluon dynamics. In this study, the electromagnetic properties of the theoretically predicted singly-charmed state, $\mathrm{X_{AV}}$, with the quantum numbers $\mathrm{J^P = 1^+}$ is investigated within the framework of the QCD light-cone sum rules method by considering the diquark-antidiquark configuration of this state. 

From the results obtained, it can be seen that the magnetic moment is sufficiently large to be experimentally accessible, while the quadrupole moment is obtained as small but non-zero values, indicating non-spherical charge distributions. A comparison of our predictions for the magnetic and quadrupole moments of the $\mathrm{X_{AV}}$ state with the predictions of other phenomenological models, such as lattice QCD, quark model, chiral perturbation theory,  and so on, would be interesting. The predicted results in this study for the magnetic and quadrupole moments of this state, along with the results for the mass, width, and other decay properties of this state and the comparison of the acquired results with the existing and future experimental data, can reveal the inner structure of this state.

\bibliographystyle{apsrev4-2}
\bibliography{XAVdiquark.bib}

\end{document}